# HOW (UN)HAPPINESS IMPACTS ON SOFTWARE ENGINEERS IN AGILE TEAMS?

Luís Felipe Amorim, Marcelo Marinho and Suzana Sampaio

Department of Computer Science (DC),
Federal Rural University of Pernambuco (UFRPE),Recife, Brazil

## ABSTRACT

*Information technology (IT) organizations are increasing the use of agile practices, which are based on a people-centred culture alongside the software development process. Thus, it is vital to understand the social and human factors of the individuals working in agile environments, such as happiness and unhappiness and how these factors impact this kind of environment. Therefore, five case-studies were developed inside agile projects, in a company that values innovation, aiming to identify how (un)happiness impacts software engineers in agile environments. According to the answers gathered from 67 participants through a survey, interviews and using a cross-analysis, happiness factors identified by agile teams were effective communication, motivated members, collaboration among members, proactive members, and present leaders.*

## 1. INTRODUCTION

In the twenty-first century, the cause-effect relationship between an individual's emotional state and his or her results in the work environment has been pointed to as a matter of study for researchers [1, 2, 3, 4, 5]. Most of those studies have their origin in the technology area, which is explained mainly by an essential milestone in software engineering's history: the appearance of the Agile Manifesto [6].

Issues with projects and dissatisfaction with a cumbersome process structure while developing software were the reasons for the arrival of new methodologies proposed by a group of developers. These methodologies were named "agile" and standardized to share some principles such as a cyclic development process, good communication and reduction of extensive documentation [7, 8]. Projects that follow agile principles described in Agile Manifesto are identified as ``agile projects''.

Boehm and Lyubomirsk [2] concluded that happy people are more satisfied with their jobs and more autonomous while executing their assigned tasks. They show better performance when compared to less happy individuals [2].

Meanwhile, satisfaction is not the only characteristic considered to measure progress and achieved results in a team. In technology environment, the relationship between work and other social and human factors might be much more relevant when an organization needs to reach its desired goals, such as, performance and productivity [9, 10, 5, 11, 12,13, 14, 15], quality [3, 16], social interactions between software developers [11, 17], and motivation [1, 18].

However, based on Diener's happiness definition [19] where being happy or unhappy is associated with the frequency of positive and negative experiences, the above cited factors have a strong connection and relevance with software engineers' affects in their work environment.

Concurrently with the great agile development process evolution, there is an outstanding people-centred framework where the whole work model is oriented to people, with individuals and interactions being more valued than processes and tools [6]. Therefore, the happiness of all stakeholders is considered an essential element to an organization`s success and should be related to the considered social and human factors [20]. As long as software development is an activity mostly relying on people, most results will be impacted by social and human factors as well [11].

This paper intends to answer the following question: "How does (un)happiness impact software engineers in agile environments?". In order to do that, a theoretical and practical study was executed in a company located in Porto Digital [21], Recife, Brazil. In this study, we analysed five agile projects and information provided by each one of their software engineers to identify how their happiness or unhappiness impacts the software development process.

This paper is organized as follows: in Section 2, we introduce the background of the problem and define our research questions. Section 3 describes the methods we use. Sections 4 and 5, present the results, their implications and limitations, respectively. Finally, in Section 6 we present our conclusions and future research directions.

## 2. BACKGROUND

The success of software management and development does not depend only on technologies and artefacts but also on human decisions about the process [22]. On the other hand, there is little research to understand social and human factors when compared to the strong interest and number of papers that aim to identify patterns inside communication and practical coordination [23].

Through statistical analysis and several factors connected to the work environment, it was possible to establish relevant associations between individuals' personality factors and tasks executed by them in their workplace [22]. Therefore, it is reasonable to state that focusing on people and providing incentives for them to be happy and satisfied in information technology (IT) workplaces implies better productivity and software quality [10].

Dozens of articles and theoretical studies were analysed, aiming at a better understanding of the relationship between happiness and productivity, and the conclusions show that happy people are more productive [14]. Indicators used to measure happiness were, for example, satisfaction at work, work environment quality, self-life satisfaction, positive and negative feelings.

Furthermore, many other studies have been conducted about social and human aspects of software engineers and work environment. Unhappiness has been linked to this environment in a survey within qualitative research that revealed 49 consequences of it on IT organizations, such as "low code quality", "low productivity", "stress", "burnout", and "frustration" [24]. Another paper focused on exploring the causes of the software engineers` frustration about their code and tasks [25].

It is also known that researches related to software development teams' performance may help the understanding of an individual's motivation to execute specific tasks properly. Therefore, development process becomes more incisive when all members' well-being is supported and a consequent sustainable culture inside software projects is created [11].

Graziotin et al. [10] performed some studies based on the Scale of Positive and Negative Experience (SPANE) [19] in order to identify emotions of 42 computer science students and established a secure connection between their analytical capacity and their creativity [10]. The need for promoting discussion about human aspects was also identified in one of their studies as an essential matter when dealing with software engineers [5].

Building upon the previous findings and ideas, another study was conducted by Graziotin et al. [26] in which a discussion based on happiness and unhappiness was the centre of a qualitative study through a survey of 317 respondents. As a result, 42 consequences of unhappiness and 32 consequences of happiness were identified [26]. Furthermore, it is stated in this research that maximizing happiness may be achieved by maximizing positive experiences or by minimizing the negative ones.

However, the researches cited above were executed in general environments. Van Kelle, E. et al. [27] developed a vital study that connected social and human factors and agile projects in Netherlands. In short, the authors built a conceptual model about social factors that might influence success in agile projects and also executed a test of 40 projects in 19 German organizations. The main purpose was to identify if the team size is a determinant factor in an agile project's success. The outcomes indicate that value of congruence, degree of adaptation and leadership were factors identified as essential while the team size premise proved to be wrong. Nevertheless, studies with a specified focus on agile development environments were not found in Brazil. Therefore, by having emotions and feelings linked mainly to software engineers` performance and cognitive process [9] and an environment in which adaptation and quick answers to changes is often required [28], it would be useful to understand how (un)happiness impacts software engineers in agile environments in an agile-driven company.

Based on the literature review, it is possible to affirm that inside agile projects, where the focus is on people over tools [6], happier people produce more and better [14], while those less happy have their performance negatively affected and are not able to deliver results with decent quality. Then, this article aims to answer the following question: *How does (un)happiness impact software engineers in agile environments?*

## 3. METHOD

This paper aims to understand the relationship between a participant's (un)happiness and social and human factors linked to each team. It is also classified as a case study, according to Runneson and Höst [29] and uses a qualitative approach based on Seaman [30] and Yin [31]. As an essential tool, a survey was created and used.

Looking to verify and answer this paper's research question, a big company in Recife that bases its everyday work on innovation culture and has the most projects in the most significant areas - telecommunication, commercial automation, finances, media, energy, health and agribusiness, was chosen.

This company's head office is located on Porto Digital [21], in Recife, Brazil. It was created in 1996 and has more than 500 employees who work in the distributed facilities in the cities of Curitiba, Sorocaba, and Manaus, besides Recife.

The study was conducted inside this company with 67 participants from five different agile projects (see Table 1). Each agile project was considered a distinct unit of study, based on the literature about agile principles and values [6, 32, 33, 34, 35, 36].

Table 1.  Case study teams characteristics.

| Team | Team Size | Type of Software | Distribution |
|------|-----------|------------------|--------------|
| *Alpha* | 6 participants | Mobile software testing | Global |
| *Beta* | 10 participants | Mobile development | Global |
| *Gamma* | 9 participants | Mobile development | Global |
| *Delta* | 26 participants | Web application development | National |
| *Epsilon* | 16 participants | Web application development | Global |

We observed the projects from July 2019 to December 2019. Specifically, one of the authors observed team's ceremonies, including daily stand-ups, sprint planning, backlog grooming, and sprint retrospectives. The same author also conducted surveys and interviews with each member of the project, which were recorded and transcribed. The interviews took approximately one hour and followed an interview protocol available from *https://bit.ly/2Ocxuev*. Besides, a survey was used as a research tool and can be found at *https://bit.ly/36CjG3f*.

To conduct the interviews and surveys, we initially performed a literature review. The data from the literature review were labelled through qualitative coding (open coding) to identify similarities and sort them to describe the constructs related to (un)happiness. Charmaz [37] clarifies that coding means that we attach labels to segments of data that depict what each segment is about.

In sequence, the constant comparative method of qualitative analysis [38] was adopted to compare each code from the same paper and those from other papers. As we continuously compared the codes, many fresh concepts emerged. As the process of data analysis progressed, relationships among categories and memos were recorded to keep us involved with the analysis, thus helping to increase the level of abstraction of our ideas about the codes, concepts, categories, and possibly even relationships.

Thus, we extracted categories to elaborate the questions. These categories can be seen in Tables 2 and 3

Table 2.  Categories related to happiness.

| Positive categories | References |
|---------------------|-----------|
| Productivity | [39, 18, 26, 5, 4] |
| Motivation | [1, 18, 26, 40] |
| Satisfaction | [41, 18, 14] |
| Creativity | [26, 10, 18] |
| Collaboration | [11, 18] |
| Engagement | [11, 26] |
| Efficiency | [41] |
| Salary | [41] |
| Time | [41] |
| Adaptation | [11] |
| Empathy | [11] |
| Problem-solving ability | [10] |
| Focus | [18] |
| Professionalism | [18] |
| Stability | [18] |
| Cognitive performance | [10] |
| Perseverance | [26] |
| Self-confidence | [26] |
| Pride | [26] |

Table 3. Categories related to unhappiness

| Negative categories | References |
|---|---|
| Demotivation | [18, 26, 40, 24] |
| Low productivity | [12, 39, 24, 26] |
| Frustration | [26, 25, 42] |
| Anxiety | [24, 42] |
| Rage | [19, 24] |
| Low self-esteem | [19, 24] |
| Activities delay | [24, 26] |
| Process deviation | [24, 26] |
| Dissatisfaction | [22, 24] |
| Low problem-solving ability | [10] |
| Monotony | [42] |
| Poor cognitive performance | [24] |

Besides, in order to do the survey, the SPANE from Diener [19] was adapted and used to understand the (un)happiness perception of all software engineers in the agile projects. In SPANE, Diener [19] elicited 12 feelings around the human experience, 6 positive and 6 negative. Then, each person was invited to evaluate their feelings along a certain period using the Likert scale from 1 to 5 for each item in which 1 means "Never" and 5 means "Always".

We performed a literature review and mapped the 8 most frequent feelings (4 positives and 4 negatives) that impact software engineers as reported by the papers we analysed and added them to the ones present in SPANE [19]. All other aspects of SPANE and the methodology were preserved. At the end of this mapping we had 10 positive feelings and 10 negative ones (see Table 4).

Table 4. Positive and Negatives feelings based on SPANE and other articles.

| Positive feelings | References | Negative feelings | References |
|---|---|---|---|
| Positive | [19] | Angry | [19] |
| Good | [19] | Negative | [19] |
| Pleasant | [19] | Bad | [19] |
| Happy | [19] | Sad | [19] |
| Joyful | [19] | Unpleasant | [19] |
| Contented | [19] | Afraid | [19] |
| Motivated | [11, 1, 40] | Bored | [42] |
| Engaged | [26, 11, 18] | Anxious | [42, 24] |
| Productive | [14, 9, 18] | Frustrated | [25, 26] |
| Collaborative | [11, 18] | Feeling of delay | [24, 26] |

The case study method is well established in software engineering [29] and this paper applied a cross-analysis to explore similarities and differences among cases [43]. We used several cases to establish general amplitude and conditions of applicability in each one of them. Results and context are displayed in Section 4.

Finally, we employed a cross-case analysis to triangulate our findings.

## 4. RESULTS

In this section, all five case-studies and their members are introduced and compared between one another based on the definition from Miles et al. [43].

### 4.1. Within-Case Findings and Analysis

After understanding software engineers' (un)happiness perception in each case study and their vision about its impact inside the agile projects, we made another set of questions based on social and human factors mentioned in Tables 2 and 3. These questions were created to understand the reality of processes and tasks in the projects and also to get a clear picture of the members' mutual relationships from them. To evaluate all the responses, a Likert scale was used in which 1 meant "Totally disagree" and 5 meant "Fully agree". Table 5 shows the participants' information.

Table 5.  Case study participants

| Team | Participant roles | Participant age | Participant experience |
|------|-------------------|-----------------|------------------------|
| *Alpha* | Tester (5)<br>Project Lead (1) | Between 20 and 25 (4)<br>Between 26 and 30 (2) | Less than 1 year (3)<br>1 - 5 years (3) |
| *Beta* | Developer (10) | Between 20 and 25 (3)<br>Between 26 and 30 (4)<br>Between 31 and 35 (3) | Less than 1 year (2)<br>1 - 5 years (4)<br>10 - 15 years (3)<br>More than 15 years (1) |
| *Gamma* | Technical Lead (1)<br>Developer (8) | Between 20 and 25 (7)<br>Between 26 and 30 (1)<br>More than 35 (1) | Less than 1 year (3)<br>1 - 5 years (4)<br>6-10 years (1)<br>More than 15 years (1) |
| *Delta* | Developer (12)<br>Project Manager (1)<br>Technical Lead (2)<br>Tester (6)<br>UX/UI Design (5) | Between 20 and 25 (6)<br>Between 26 and 30 (9)<br>Between 31 and 35 (4)<br>More than 35 (7) | Less than 1 year (2)<br>1 - 5 years (9)<br>6-10 years (8)<br>10 - 15 years (6)<br>More than 15 years (1) |
| *Epsilon* | Developer (9)<br>Project Manager (1)<br>Technical Lead (1)<br>Tester (4)<br>UX/UI Design (1) | Between 20 and 25 (1)<br>Between 26 and 30 (6)<br>Between 31 and 35 (4)<br>More than 35 (5) | 0 - 5 years (2)<br>6-10 years (9)<br>10 - 15 years (4)<br>More than 15 years (1) |

#### 4.1.1. Case 1: *Alpha*

The first case analysed is a small project with only 6 people as can be seen in Table 1. It is software distributed worldwide and its members are 2 women and 4 men between the ages of 20 and 25 (see Table 5).

Although the team has a young profile, it has been working for 5 years and there is no finish date estimated yet. Therefore, its customer satisfaction is high, which confirms the vision of Serrador et al. [41], who reported that agile methodologies have a direct impact on a project`s success. Results are presented in Figure 1.

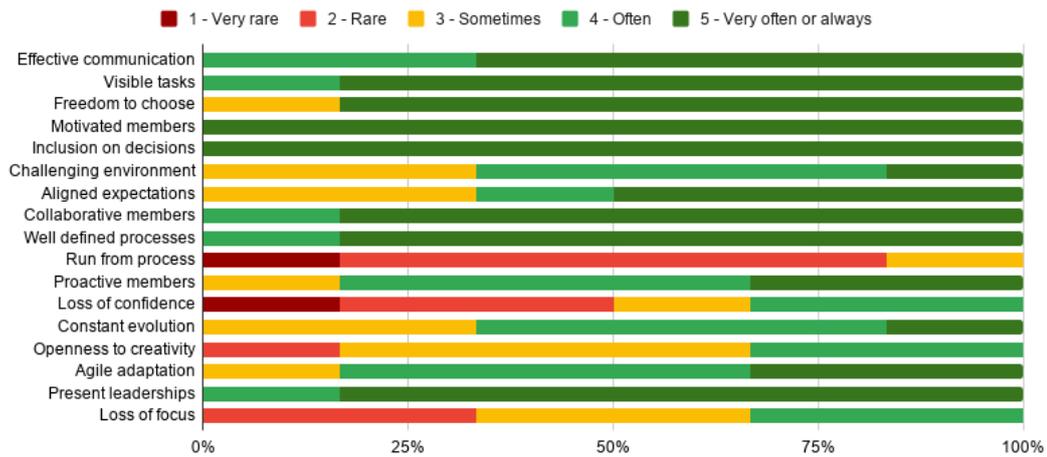

Figure 1. Results of team factors - *Alpha* case study

### 4.1.2. Case 2: *Beta*

The second study was done in an auto-managed agile team of 10 people that has been working for 1.5 years with no due date. There are software engineers of many different ages as can be seen in Table 5).

As there is no manager on this project, the organization of activities is done based on formal communication among all members through *kanban* boards. Such board is considered by Destefanis et al. [16] as the central idea of communication inside agile philosophy. The members also communicate informally during the team's daily process. Van Kelle et al. [27] state that such team organization creates shared values helps with trust and promotes strong interpersonal relationships.

Results from the team`s perspective are found in Figure 2

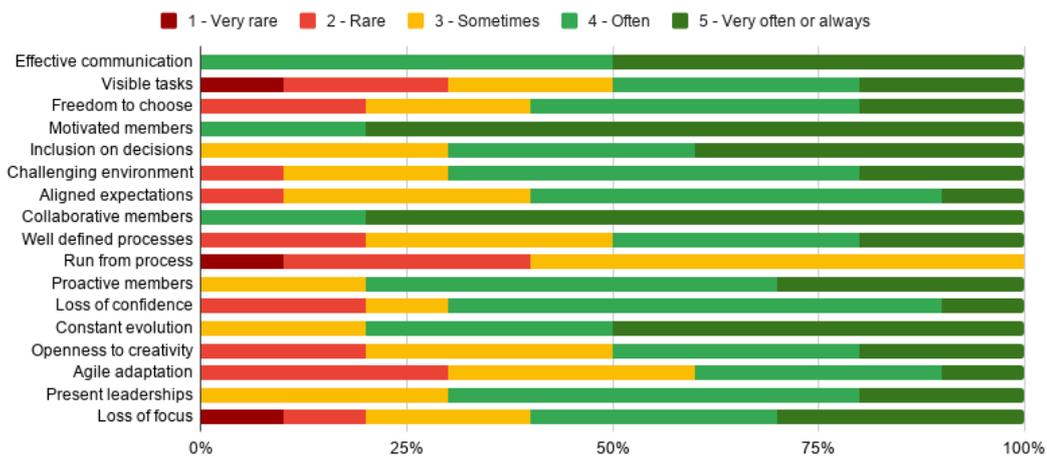

Figure 2. Results of team factors - *Beta* case study

### 4.1.3. Case 3: *Gamma*

This project, distributed worldwide, is also shown as an agile one with 9 software engineers. From these, 2 are women and 7 men. All of them are software developers. It has already achieved 5 years and has no due date. Most of its members' ages are between 20 and 25.

Results from the team's perspective are found in Figure 3.

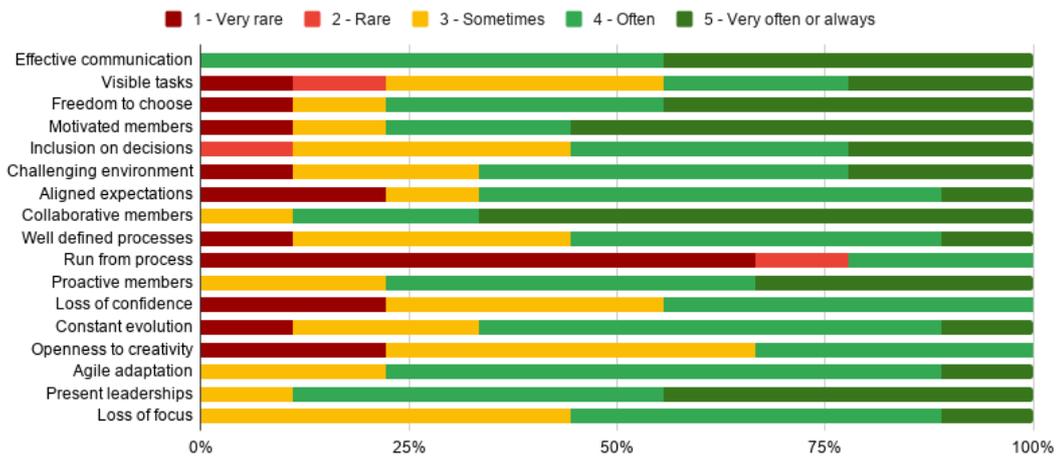

Figure 3.  Results of team factors - *Gamma* case study

### 4.1.4. Case 4: *Delta*

The case number 4 consists of a subdivided agile project, which has existed for the last 7 years. Nowadays, Twenty-eight software engineers are present, and they work on Web application development for an international client. Nine of them are women and seventeen are men, from 25 to 35 years old.

Results from the team's perspective are found in Figure 4.

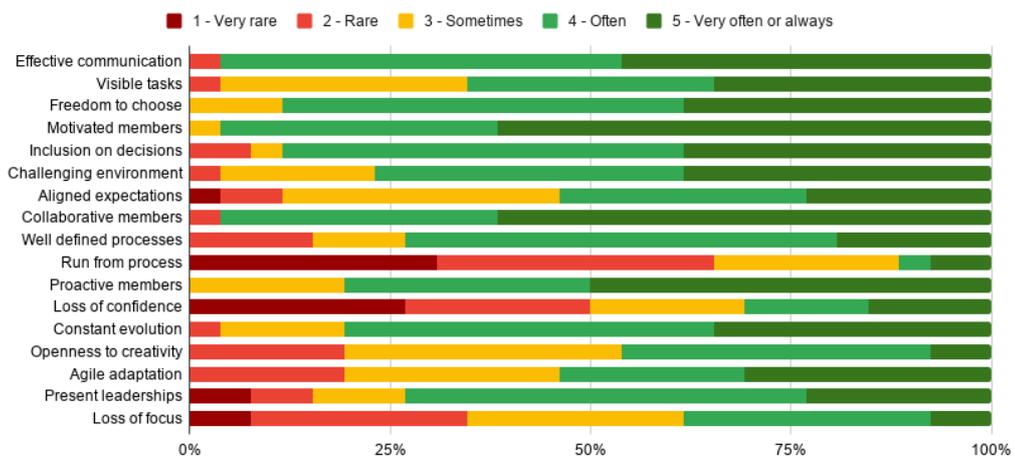

Figure 4.  Results of team factors - *Delta* case study

### 4.1.5. Case 5: *Epsilon*

This case study is of an agile project existent for the last 2 years with 29 software engineers. From these members, we gathered 16 responses, all of them from men between the ages of 20 and 35. Most of them have more than 6 years of experience, which brings an experienced profile to this team.

Results from the team's perspective are found in Figure 5.

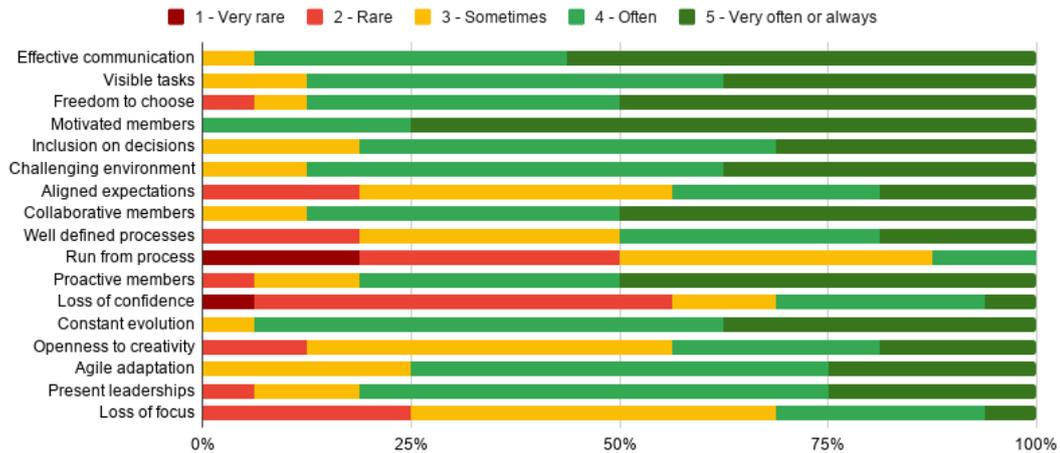

Figure 5.  Results of team factors - *Epsilon* case study

## 4.2 Cross-Case Analysis

### 4.2.1. Happiness Perceptions

To comparatively analyse perceptions of (un)happiness between the studied projects, those software engineers who were identified as the happiest from each project were isolated using the results collected from the adapted SPANE (Figure 6).

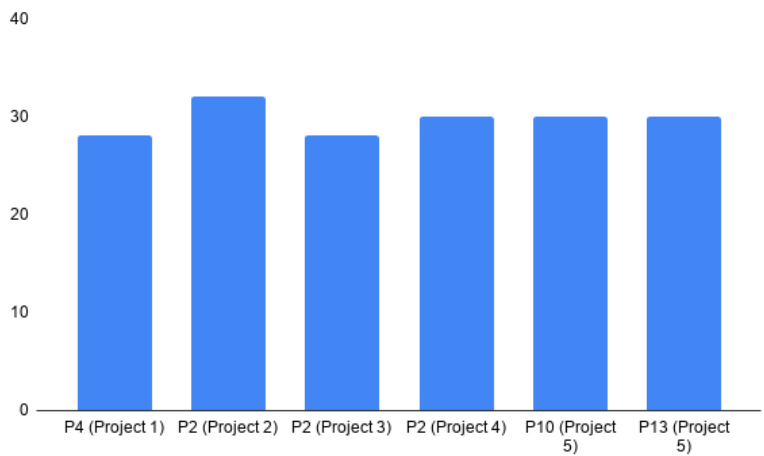

Figure 6.  Software engineer happiness perception

The engineer with the highest happiness index in the *Alpha* case study is a woman between 20 and 25 years of age, with less than 1 year of experience. A graduate student who performs the role of a tester in the project in which she is involved. According to her responses on the scale, widespread factors contributing to her happiness were collaboration, communication, satisfaction, engagement and high productivity. On the other hand, she was anxious and a little insecure.

In the *Beta* case study, the happiest software engineer is a 30 to 35-year-old male with a master's degree and over 15 years of work experience working as a developer. According to him, the factors that most impacted his positive results were high productivity, collaboration, communication, satisfaction and confidence in his skills that were present in the agile environment.

In the *Gamma* case study, a man between the ages of 20 and 25, still an undergraduate student with just over 1 year of experience working as a developer, was considered the happiest and showed high engagement, motivation, communication and collaboration as positive results of his performance within the agile team. However, he felt a little insecure and anxious.

In the *Delta* case study, a woman between 20 and 25 years of age and still an undergraduate student has been identified as the person with the highest perception of happiness within her team, where she acts as a tester. She has just over 1 year of experience in the job market and has highlighted positive points regarding her performance, such as high productivity, engagement, motivation, satisfaction and collaboration. Among the negative points, anxiety and insecurity were more evident.

In the *Epsilon* case study, two men appeared tied on the scale used for the perception of happiness and unhappiness. Both between 25 and 30 years old and with more than 6 years of experience, but in different roles. The first is a designer who is very engaged, satisfied, communicative and collaborative, but feels a little bored. The second is a developer of the same project who identified himself as very productive, engaged, motivated, satisfied and collaborative. However, feelings of slowness, insecurity and anxiety are evidenced as negative points.

By analysing each of these individuals, it is possible to find some commonalities in the agile environments they are working in, such as high satisfaction and collaboration. These characteristics are expected as factors in happy software engineers in general development environments [18] and specifically in agile environments as the whole dynamic is people-oriented [6]. On the other hand, it is possible to see that younger people tend to report a little more anxiety and insecurity, even when they appear to be happy and engaged. Additionally, developers seem to have a more palpable sense of backwardness compared to testers and designers.

### 4.2.2. Unhappiness Perceptions

Similarly to the previous subsection 4.2.1, this subsection provides an overview of the least happy individuals for each case study, according to the data and responses collected through our instrument. Figure 7 displays these data.

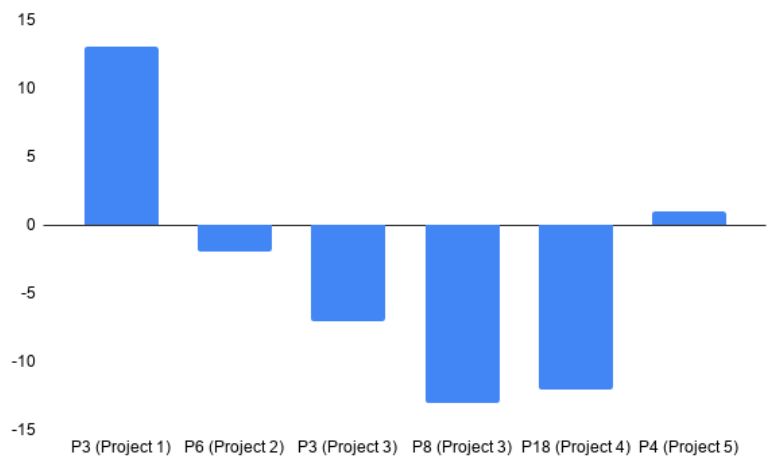

Figure 7. Software engineer unhappiness perception

According to Figure 7, it is possible to see that in the *Alpha* case study, the software engineer presented is not unhappy, but was considered the least happy within his project. He is a man between 20 and 25 years old with less than 1 year of experience, graduate student and performing the role of a tester. In the analysis of the concepts attributed by him, it is possible to identify a neutral pattern where there is no insecurity or anxiety, but there is a lot of boredom, dissatisfaction and a loss of confidence and motivation. Still, he claims to have good productivity and good communication with his teammates.

In the *Beta* case study, it can see a man between 20 and 25 years old, still graduate student and with experience between 1 and 5 years working as a developer. In his analysis, we found high frustration, dissatisfaction and boredom plus some insecurity. Although these negative points are evident, he has good communication with the team.

In the *Gamma* case study, two software engineers were identified with a noticeable perception of unhappiness. The first is a woman between 20 and 25 years old, an undergraduate student with less than 1 year of experience. She works as a developer on the team and presents, according to the analysis, critical points such as insecurity, frustration and low self-confidence. She reported, "not feeling stimulated to finish an activity". Communication with the team is good according to her viewpoint, but her collaboration with the team is not noticeable.

The second unhappy engineer in the *Gamma* case study is also a 20 to 25-year-old male graduate student with less than 1 year of experience in the job market. He acts as a developer and shows a high rate of frustration, low communication and team collaboration, boredom, low self-confidence and high dissatisfaction. He even reported "being unmotivated".

In the *Delta* case study, a woman with a graduate degree, designer, between 20 and 25 years old, with less than 1 year of experience is the least happy person. She can be considered a sad, negative, bored and very anxious person. On the other hand, according to the concepts attributed by her to each feeling present on the scale, there is not good communication or collaboration within her team. Productivity is also weak.

In the *Epsilon* case study, as in the first case, there is not an unhappy software engineer, but a person with a lesser sense of happiness within his team. In this case, a male, 35-year-old developer, graduate degree, with 10 to 15-years of experience in the job market thinks his productivity, communication, and collaboration are good, but there is also frustration, dissatisfaction, and boredom with the tasks he performs.

Thus, it is possible to find a pattern by analysing the data from these members among themselves. All those listed here as unhappy software engineers, except for the last one, are people between the ages of 20 and 25 and with less or a little over 1 year of experience. All show signs of anxiety and dissatisfaction with their roles and performance within their respective teams. In the *Epsilon* case, in particular, it is assumed that due to his age and the number of years of work, the individual is unmotivated because of the monotony of his activities.

### 4.2.3. (Un)Happiness in agile environments

In this section, the perception of each team, in general, will be analysed. The analysis will be performed based on the factors of each team, generated by the concepts attributed to each of the 17 statements present in the survey protocol *https://bit.ly/36CjG3f.*

To facilitate this step, the concepts "Very rare or never" and "Rarely" have been unified as well as "Very often or always" and "Often" and represent negative and positive side, respectively. The concept "Sometimes" will be considered as neutral.

It is possible to verify in the factors of the *Alpha* case study (Figure 1) that the communication between members happens only positively and it has a positive relationship in the opinion of all members. Activities are visible and information always available.

In the *Beta* case study (Figure 2), even though everyone agrees on a relationship between capable team communication and happiness, only 50% of participants confirm the same positive relationship between the visibility of project activities and the ease of communication.

In the *Gamma* case study (Figure 3), the percentage of members who confirm that the visibility of activities influences communication success is slightly lower (44%) compared to the results of project *Beta*. The members also agree that effective communication occurs within the team and positively influences team performance.

In the case study of the *Delta* team, Figure 4 displays a directly proportional relationship between effective communication of the members and their performance. However, this result is not absolute, since 4% of the participants denied this relationship. The same portion of participants also denies an express relationship between ease of communication and performance, while 31% remain neutral and the remaining 65% confirm its presence.

Finally, the results of the *Epsilon* case study (Figure 5) show that only 6% of the participants do not see the direct influence of effective communication among colleagues on happiness and consequent team performance, while 94% claim the veracity of this relationship. A small number, 13%, disagree that having activities be visible facilitates communication and leads to a happier environment and better performance. On the other hand, 77% confirm this relationship.

By inspecting the motivation part of the members of the five cases studied, it is possible to see that in all of them, the vast majority of members understand the importance of the tasks and are free to choose and learn during their execution. They also agree that this freedom motivates and positively affects their performance. Only in projects *Beta*, *Gamma*, and *Epsilon* were there software engineers who do not agree, with the proportions being 20%, 6% and 6%, respectively. Project members from *Alpha*, *Beta* and *Epsilon* unanimously affirm that they carry out their tasks with greater commitment when motivated. While 78% and 62% project members from *Gamma* and *Delta*, respectively, agree with this statement.

Regarding the inclusion of all members in the decisions and its impact on motivation and better team performance, only project *Alpha* unanimously confirmed the occurrence and importance of the fact. Most of the members of the other projects also agreed, but some responded neutrally or disagreed.

When asked if the environment could be characterized as challenging and whether this had a positive impact on the team performance, most participants in each project agreed. Although all projects had some members who responded neutrally, only projects *Beta*, *Gamma*, and *Delta* showed a small number of members respond negatively.

Regarding the alignment of expectations with their customers and the involvement of the entire team in this process, the first three projects (Figures 1, 2 and 3) set themselves up as engaging and enjoyable environments for their members. In the fourth case study, the project was balanced between neutrality and confirmation, which may be a warning for the project's sustainability, as a collaboration with the client is more important than contract negotiation manifest. However, in the analysis of the last case study, only 44% of respondents note the involvement of members in aligning expectations with the client.

Similar to customer collaboration, member interaction is also present as a value in the Agile Manifesto [6]. Based on this, a question was asked to verify that each project was indeed collaborative, with members available and open to assist each other, and whether this had a positive impact on team performance. Projects *Alpha* and *Beta* unanimously confirmed this factor in their environments and stated that helping others positively impacts performance. Project *Delta*, in turn, was the only one that received a negative concept share (4%). Case studies *Gamma* and *Epsilon* were mostly rated as positive in this regard but also obtained a neutral rating from the evaluations of some members that totalled 11% and 13%, respectively.

According to Zelenski et al. [14], happy people are more productive people. Therefore, some questions related to productivity were asked. In the first question, the participants responded whether the processes performed by their team were well-defined. Project *Alpha* was the only one with 100% positive feedback for the presence of well-defined processes and their impact on team organization and productivity. The members of other projects had different opinions. Studies *Beta* and *Epsilon* report that only 50% of participants agree on the existence of well-defined processes within their teams. In projects *Gamma* and *Delta*, the results were 54% and 73%, respectively.

Based on another value of the Agile Manifesto [6], that responding to change should be considered more important than following a plan, a question about process flight coupled with frustration and productivity was asked. Responses from all the projects, with the exception of project *Beta*, show that they continue to use the same processes even if they do not have the desired effects without looking for alternative paths. This is a matter of concern and attention since all the projects studied are considered agile according to the references used in this research. Furthermore, most participants in all studies guarantee better choices for code maintenance when they feel happy. They also often seek out new activities or help other colleagues after their tasks have been completed. Thus, they experience a sense of usefulness and importance within each team.

In the analysis, project *Beta* was the only one identified as an environment where most members, 70%, feel they are distressed when they fail to meet all the expectations and goals set for a stage and lose confidence in their skills. All the other projects did not have this feature expressly identified by most of their members.

On the other hand, all studies presented most individuals as showing a constant evolution of knowledge and affirming the positive impact of this on their performance. The biggest highlight was study *Epsilon* where 94% of members denoted this perception.

All studies are lacking in the openness of members to generate new ideas. Project *Delta* was identified as the worst among all, where only 56% of software engineers identified this openness and confirmed its relationship to each individual's happiness and well-being.

A worrying scenario is the recognition of agile methodologies by the teams, since only projects *Gamma*, with 78%, and *Epsilon*, with 75%, were aware that they are inserted in an environment with agile practices and understand each stage of development and activities carried out. Projects *Gamma*, *Beta*, and *Delta* do not seem to recognize the environment in which they are inserted, as they have only 33%, 40% and 54%, respectively.

Regarding the leaders, all of them are very present and accessible in all projects. However, project *Alpha* and *Gamma* leaders seem to have a more significant influence on the day-to-day life of the members, as they were identified as 100% willing to help and motivate their teams during the development process, recognizing achievements and successes and impacting on the happiness and motivation of individuals.

The engineers' focus, as the last point, was analysed among all teams regarding happiness and unhappiness. Project *Beta* presented itself as an environment where members often keep their attention on activities even if they become complex and laborious. However, if there is a delay in solving the problems encountered, stress and feelings of inability begin to become evident. Concerning the other projects, all presented divergences in the three concepts when analysed through the overview.

## 5. DISCUSSION

According to the data found and analysed in this article, it is possible to make some considerations about the guiding question: "*How does (un)happiness impact software engineers in agile environments?*" defined at the beginning of this paper.

According to the values present in the main source of agile methodologies [6], interaction between people is a fundamental point in environments that use them through a formal [16] or informal way [27]. Given this, Van Kelle [27] points to communication and leadership style as success factors in an agile project and also states that lack of effective communication or the existence of misunderstandings is the main reasons for the failure of a project.

Complementarily, Fagerholm et al. [11] indicated in their studies that there is an intrinsic relationship between the ease of communication established within teams and the happiness or unhappiness of the members; they stated that creating a communicative atmosphere implies benefits for the performance of individuals. In the present study it was possible to confirm the influence of good communication on team members' happiness and consequent good team performance, as can be seen in the cross-analysis present in section 4.2 where the "effective communication" factor was addressed.

Destefanis et al. [16] also indicate the importance of task visibility through boards available to all. In their view, this element represents the central vision of communication in agile environments and is considered a facilitator in the communication and organization of teams. Thus, it is possible to note that members of most projects are adept at using this technique, except some members of projects *Beta* and *Gamma* who do not seem to see the importance to the same extent as the other project members. Therefore, there is a small risk of failure in these projects according to the study of Van Kelle et al. [27], which points out that most projects do not fail due to technology, but because of social and organizational problems, lack of (effective) communication and misaligned teams.

Besides, there is a possibility that the definitions of the objectives may not be apparent to everyone, which negatively affects the motivation of individuals [40]. However, unmotivated members are an essential consequence of unhappiness for software engineers [26], which implies low performance.

On the other hand, [1] claim that agile environments lead to greater motivation among software engineers and high motivation is considered an indication of happiness [26]. Therefore, in the present study, it is noted that all projects present the majority of their members happy and motivated, which can be verified by the scales used for the perception of happiness individually and from the factors conceptualized by each team about the factor in question.

França et al. [40] suggest that a motivated person is also engaged and focused. When analysing the teams, the inclusion and engagement of all members in the definitions of activities and objectives happens and reflects well-being according to most participants. This can be seen in Figure 1, as no member of this project was considered unhappy and everyone claims to be engaged and included in the decisions. The same cannot be seen when analysing Project *Gamma*, where two members were identified as unhappy and report immense demotivation and, according to [40], consequent unhappiness.

Additionally, [18] claim that happy software engineers are the ones who are satisfied. Satisfaction is associated, according to [22], with the ability to make decisions and with the level of challenge found within the work environment. In this sense, [18] mention that job satisfaction affects physical and mental health, which is consistent with the work of [14] who associated factors like these to higher productivity and better performance. Thus, *Gamma* case study is the only one that suggests a less inclusive and satisfactory decision making for its participants. On the other hand, according to most respondents, all projects present challenging environments.

With a large number of challenges and the need for interaction among members [6], teams related to software development should be engaged and focused on both negotiating and aligning expectations with the customer and in a constant discussion and improvement of internal performance [11]. In their study, [26] found a relationship of engagement and perseverance with individual's happiness within a software development process.

Therefore, in this work, it is possible to identify more collaboration among individuals working in projects with a higher rate of happy people.

It was also noted that 4 of the 5 teams do not appear to have well-defined processes, as reported by their respective members. All of them also claim they do not deviate from existing processes and they do not take new actions when some stage does not show satisfactory results. This violates one of the principles present in the Agile Manifesto [6]: "Agile processes harness change for the customer's competitive advantage", which can affect happiness, since [26] mention that being happy or unhappy [42], is also associated with consequences arising from the processes established in software development.

Furthermore, adapting to new environments and technologies is considered an essential requirement for developers, according to [25], who claim that there is a feeling of frustration when difficulties are not fully faced and overcome, causing a negative impact on performance. Although most members claim not being frustrated in this regard and are continually evolving, learning new technologies, adapting to new environments and tools, and overcoming challenges, the frustration factor of not achieving goals in an expected period was found in the *Beta* case study, where 70% of members reveal that they feel technically inadequate when they are unable to reach all the objectives established for a given stage of development.

Related to this, it was possible to identify the lack of openness to creativity within all study units, contradicting the findings of [26] who say there is a strong relationship between high creativity and happiness and also state that its reduction should be considered problematic in various situations. In the present study within an innovation company, the absence of this element from the fundamental factors of happiness can be a sign of an internal failure of the projects in question. This opens the possibility for further investigations with greater depth and attention in future works.

Finally, when analysing the results of this research related to the factors of leadership, degree of adaptation and focus mentioned as essential in the study by [27], it is possible to identify some interesting points. The recognition that comes from the leaders and stakeholders for the members of their respective teams is an essential factor linked to happiness, as can be seen in the reports of some participants: "Recognition from the client positively affected my performance", "Having my job and recognized technical evolution, which made me work with more commitment". The degree of adaptation to agile procedures is also an essential factor to be considered since some teams do not understand the environment in which they operate very well. Only 2 of the 5 projects analysed confirmed their awareness of the stages and their contribution to the development process. Therefore, it is necessary to evaluate more carefully the communication of agile teams, since this factor is presented as essential to improving the performance of an agile team [27]. Through this research and based on the answers obtained, it was not possible to establish the relationship between focus and happiness or unhappiness.

## 5.1. Limitation

In this multiple-case study, the uniqueness of corporate, team and project factors of each case render valid comparison and theoretical generalisation more difficult. Specifically, in data gathering, a limitation of the interview technique is that the interviewers may introduce bias by their approach to the interview, word emphasis, tone of voice, body language, and question rephrasing [44]. While the interviewers must be aware of their effect on the data being gathered, this limitation is accepted as inevitable by interpretivists.

However, the study also limited its scope to studying only individuals within agile projects in a single technology company, which may not reflect the real state of the relationship between happiness and unhappiness within agile environments, since there are several such environments and several companies that use agile practices in their projects.

The psychological study of an individual is also limited as the only basis for the results were the individual perceptions of each participant.

Finally, the relationship of high creativity and happiness was not conclusive within the study units used, which can be the subject of a more detailed study.

## 6. CONCLUSIONS

Creating an agile and happy environment implies performance benefits for software engineers. We observed the relationship between (un) happiness of software engineers in five projects. For this, we conducted a theoretical and practical study in a company located in Porto Digital [21], Recife, Brazil, in which we analysed five agile projects and information provided by each of their software engineers.

We have identified some happiness factors that impact software engineers in agile environments, such as, effective communication, motivated members, collaboration among members, proactive members, present leaders. However, we have not identified creativity as a factor that impacts the agile projects being pursued.

For future work, we indicate the need to replicate the present study in other organizations in order to offer a more comprehensive perspective on the topic. Additionally, a statistical analysis of the correlation between the factors related to happiness and unhappiness should be performed, so that there is an objective vision about which negative points should be minimized and which positive points should be maximized. These can then be used to provide recommendations for agile, happy, and productive environments. Also, a study to deeply verify the relationship between creativity and happiness within agile environments should be conducted.